\documentclass[useAMS,usenatbib]{mnras}

\usepackage{bm}
\usepackage{natbib}
\usepackage{graphicx}
\usepackage{amssymb, amsmath}
\usepackage[normalem]{ulem}
\usepackage[dvipsnames]{xcolor}

\title[Heating of the upper layers of the inner crust]{Accreting neutron stars: heating of the upper layers of the inner crust}
\author[N.N.\ Shchechilin, M.E.\ Gusakov, A.I.\ Chugunov]
{N.N.\ Shchechilin,\thanks{nicknicklas@mail.ru} M.E.\ Gusakov, A.I.\ Chugunov\\
Ioffe Institute, Polytekhnicheskaya 26, 194021 Saint Petersburg, Russia}
\begin{document}
	
	\date{Accepted 2022 xxxx. Received 2022 xxxx;		in original form 2022 xxxx}
	
	\pagerange{\pageref{firstpage}--\pageref{lastpage}}
	\pubyear{2022}
	
	\maketitle
	
	\label{firstpage}

%%%%%%%%%%%%%%%%%%%%%%%%%%%%%%%%%%%%%%%%%%%%%%%%%%%%%%%%%%%%%%%%%%%%%%%%%%%%%%
\begin{abstract}
Neutron stars in low-mass X-ray binaries are thought to be heated up by accretion-induced  exothermic nuclear reactions in the crust.  The energy release and the location of the heating sources are important ingredients of the thermal evolution models. Here we present thermodynamically consistent calculations of the energy release in three zones of the stellar crust: at the outer-inner crust interface, in the upper layers of the inner crust (up to the density $\rho \leq 2\times 10^{12}$\,g\,cm$^{-3}$), and in the underlying crustal layers. We consider three representative models of thermonuclear ashes (Superburst, Extreme rp, and Kepler ashes). The energy release in each zone is parametrized by the pressure at the outer-inner crust interface, which encodes all uncertainties related to the physics of the deepest inner-crust layers. Our calculations allow us, in particular, to set new lower limits on the net energy release (per accreted baryon): $Q\gtrsim0.28$\,MeV for Extreme rp ashes and $Q\gtrsim0.43-0.51$\,MeV for Superburst and Kepler ashes.
\end{abstract}
%%%%%%%%%%%%%%%%%%%%%%%%%%%%%%%%%%%%%%%%%%%%%%%%%%%%%%%%%%%%%%%%%%%%%%%%%%%%%%


%%%%%%%%%%%%%%%%%%%%%%%%%%%%%%%%%%%%%%%%%%%%%%%%%%%%%%%%%%%%%%%%%%%%%%%%%%%%%%
\section{Introduction}
%%%%%%%%%%%%%%%%%%%%%%%%%%%%%%%%%%%%%%%%%%%%%%%%%%%%%%%%%%%%%%%%%%%%%%%%%%%%%%

Accreting neutron stars (NSs) in low-mass X-ray binaries (LMXBs) are observed and actively studied for more than 50 years (\citealt*{Giacconi62,Shklovsky67,Fujimoto_ea84, ylpgc04,hpy07,heinke_et_al_09,wdp17,pcc19,Liu_ea21,pc21,Fortin_ea21}). During this time, a generally accepted picture of the basic physical processes in such systems has emerged. It includes mass transfer from the less compact binary companion (\citealt{FKD02_accretion}),  stationary or non-stationary accretion onto the NS surface (\citealt{Lasota01}), thermonuclear burning in the outermost layers of NS envelope (e.g., \citealt{johnston20} and references therein), and, finally, the burial of thermonuclear ashes by the newly-fallen material, which is accompanied by compression of these ashes by the increasing pressure (see, for example, {\citealt{Sato79}). These phenomena manifest themselves in observations, providing thus a possibility not only to estimate the masses and radii of accreting NSs (e.g., \citealt*{spw_20,Kim_ea21}), but also to extract more detailed information about the properties of superdense matter in their interiors (see \citealt{mdkse18} for a review). The reliability of such an `extraction' crucially depends on the energy release in the crust of accreting NSs during compression, which is the topic of the present Letter.
	
Thermonuclear burning in the surface NS layers converts accreted nuclei (hydrogen and/or helium) into the ashes -- a complex mixture of heavy elements, dominated by nuclei of iron or palladium group. The actual composition of the ashes depends on the accretion rate, the metalicity of the accreted matter, crust temperature, etc.
 
The ashes are gradually buried by subsequent accretion and undergo a set of nuclear reactions caused by compression. It is generally believed that these exothermic nuclear reactions heat up the crust and substantially contribute to the thermal emission of the accreting NSs during quiescence periods (\citealt{bbr98}) when the accretion is strongly suppressed. Numerous papers confront observations with the steady-state thermal models of NSs (\citealt{ylh03,hjwt07,wdp13,by15,pcc19,Liu_ea21,Fortin_ea21}, etc.) or with  calculations of thermal crust relaxation just after the accretion episode (for example, \citealt{Ruthledge_etal02_KS,syhp07,bc09,Brown_ea18,pc21}), and impose certain constraints on the properties of NS crust and superdense matter in the NS core (\citealt{Degenaar_ea21}).

However, the `traditional' models of accreting NSs (\citealt{Sato79,HZ90,HZ08,gkm08,Steiner12,Fantina_ea18,lau_ea18,SC19_MNRAS,Schatz_ea22}) have missed an important element. Namely, all of them implicitly neglect the redistribution of unbound neutrons in the inner crust. It has recently been shown that this approximation is incorrect  (\citealt{CS19_NoEquil}) and leads to violation  of the neutron hydrostatic/diffusion (nHD) equilibrium in the inner crust (\citealt{GC20_DiffEq}, hereafter GC20). In fact, the nHD equilibrium condition, given by $\mu_\mathrm{n}^{\infty}\equiv\mu_\mathrm{n} \mathrm e^{\nu/2}=\mathrm{const}$ (here $\mu_\mathrm{n}$ is the local neutron chemical potential, including the rest mass and $\mathrm e^{\nu/2}$ is the redshift factor), should be considered simultaneously with the hydrostatic TOV equations. The pioneering model implementing this condition was developed in GC20. It shows that the nHD condition alters the crust composition and the equation of state (EOS) considerably.

The subsequent work (\citealt{GC21_HeatReleaze}, hereafter GC21) has shown that the deep crustal heating $Q$, arising due to nonequilibrium nuclear reactions, decreases substantially in comparison to traditional models. This fact calls for the development of more accurate nHD models and corresponding reinterpretation of observational data on transiently accreting NSs. Here we make a next step in this direction by extending our previous work \citealt{SGC_OC21} (SGC21 below). Namely, we calculate the heat release at the outer-inner (oi) crust interface and in the upper layers of the inner crust for realistic compositions of nuclear ashes, and, making use of the thermodynamic consistency of the nHD model, determine the heat release in the remaining part of the inner crust.

%\vspace{-0.5cm}
%%%%%%%%%%%%%%%%%%%%%%%%%%%%%%%%%%%%%%%%%%%%%%%%%%%%%%%%%%%%%%%%%%%%%%%%%%%%%%
\section{Formalism}
\label{Sec:Approach}
%%%%%%%%%%%%%%%%%%%%%%%%%%%%%%%%%%%%%%%%%%%%%%%%%%%%%%%%%%%%%%%%%%%%%%%%%%%%%%

\subsection{Basic features of the nHD approach and general energetic expressions}

As it was shown in GC20 and GC21, within the nHD approach the oi interface is not associated with the neutron drip out from the accreted nuclei, as it was previously supposed (e.g., \citealt{Chamel_etal15_Drip}). Rather, it is the surface of the `neutron sea', which fills the inner crust and the core. The pressure $P_\mathrm{oi}$ at this interface can not be calculated by modelling the nuclear evolution as an initial problem. It should be determined by looking at the details of nuclear reactions in the entire crust since the actual $P_\mathrm{oi}$ value depends on the evolution of the underlying inner crust layers. $P_\mathrm{oi}$ is likely to vary at the initial stages of the accretion but settles to some definite value when the fully accreted state of the crust (FAC) is reached (GC20 and GC21).

It is instructive to start our analysis of the FAC state from consideration of a family of the crust models parametrized by $P_\mathrm{oi}$ for fixed initial composition (GC20, GC21, and SGC21). Each model from this family can be constructed by considering compression-driven nuclear evolution as an initial problem, however, at $P=P_\mathrm{oi}$  the reaction network should be switched to a procedure that accounts for the nHD state in the inner crust (see Sec.\ \ref{Sec:Reac}). It leads to explicit dependence of the considered models on $P_\mathrm{oi}$. The FAC state is a member of this family with some definite value of $P_\mathrm{oi}$. 

 In the FAC state, the crust EOS is fixed and does not change during subsequent accretion. Since accretion supplies additional nuclei to the crust, an efficient mechanism of nuclei disintegration should be active in the crust to keep EOS unchanged. As shown in GC20, GC21, the required mechanism naturally emerges if $P_\mathrm{oi}$ exceeds some threshold value. The mechanism is realized in the form of a specific instability arising in the deeper layers of the inner crust. GC21 has shown that the threshold value of $P_\mathrm{oi}$ is sensitive to the shell effects in the inner crust; it has also revealed that the actual value of $P_\mathrm{oi}$ (and the respective FAC state) can depend on the previous crust evolution. In order to accurately determine $P_\mathrm{oi}$, these effects must be carefully examined, which is a very complicated and model-dependent problem (some aspects of this problem are planned to be discussed in the extended version of GC21, which is in preparation).

Fortunately, the thermodynamic consistency of the nHD model allows us to calculate the deep crustal heating, $Q$, as a function of $P_\mathrm{oi}$ without explicitly considering the whole crust --- only EOS at $P<P_\mathrm{oi}$ is required (GC21):
\begin{equation}\label{Q_gen}
	Q=Q_\mathrm{o}(P_\mathrm{oi}) + [\mu_\mathrm{b,\, oi}(P_\mathrm{oi})-m_\mathrm{n}],
\end{equation}
where $Q_\mathrm{o}$ is the energy release in the outer crust, $\mu_\mathrm{b,\, oi}$ is the baryon chemical potential at the bottom of the outer crust, and $m_\mathrm{n}$ is the neutron mass.
Note, the deep crustal heating as seen by a distant observer is given by  $Q^{\infty}=Q\mathrm e^{\nu_\mathrm{oi}/2}$, where $e^{\nu_\mathrm{oi}/2}$ is the redshift factor at the oi interface. 

Here, as in  SGC21, we make use of the above-mentioned fact: we consider $P_\mathrm{oi}$  as a parameter. Furthermore, we assume that  $P_\mathrm{oi}$ is close to $P_\mathrm{nd}^\mathrm{(cat)}$ -- the pressure at the oi interface in the catalysed crust (see GC21 for a detailed motivation), leaving the determination of the actual value of $P_\mathrm{oi}$ for future work. In SGC21, we restrict our consideration to nuclear evolution in the outer crust and, using equation (\ref{Q_gen}), calculate deep crustal heating for multicomponent ashes (three representative compositions were considered) within the nHD approach. In this Letter, we extend the calculation to the upper layers of the inner crust, limited by the density  $\rho_\mathrm{dc} = 2\times 10^{12}$\,g\,cm$^{-3}$, 
where it is still reasonable to use the atomic mass tables (\citealt{lau_ea18,Schatz_ea22}). We explicitly calculate the heating at the oi interface, $Q_\mathrm{oi}$, and the heating in the upper layers ($\rho<\rho_\mathrm{dc}$) of the inner crust, $Q_\mathrm{i1}$. Then, using equation (\ref{Q_gen}), we determine the heating in the remaining part of the inner crust, $Q_\mathrm{i2}$:
\begin{equation}\label{eq2}
	Q_\mathrm{i2}=\mu_\mathrm{b,\,oi}(P_\mathrm{oi})-m_\mathrm n - Q_\mathrm{oi}(P_\mathrm{oi}) -Q_\mathrm{i1}(P_\mathrm{oi}). 
\end{equation}
The quantities $Q_\mathrm{oi}$, $Q_\mathrm{i1}$, and $Q_\mathrm{i2}$, combined with the outer crust heating $Q_\mathrm{o}$ calculated in SGC21, are presented below as functions of $P_\mathrm{oi}$. Direct calculations of $Q_\mathrm{oi}$ and $Q_\mathrm{i1}$ allows us to tighten the lower limit for $Q$ by requiring 
$Q_\mathrm{i2}=Q-Q_\mathrm{o}-Q_\mathrm{oi}-Q_\mathrm{i1}>0$ (in SGC21 we use the less strict condition $Q-Q_\mathrm{o}>0$). We leave a detailed discussion of the composition and reaction pathways in the accreted crust at $\rho<\rho_\mathrm{dc}$ to a subsequent publication.

%\vspace{-0.5cm}
%%%%%%%%%%%%%%%%%%%%%%%%%%%%%%%%%%%%%%%%%%%%%%%%%%%%%%%%%%%%%%%%%%%%%%%%%%
\subsection{Physics input and reaction network} \label{Sec:Reac}
%%%%%%%%%%%%%%%%%%%%%%%%%%%%%%%%%%%%%%%%%%%%%%%%%%%%%%%%%%%%%%%%%%%%%%%%%%

In SGC21, we applied up-to-date set of experimental atomic masses (\citealt{ame20}, AME20%
%
%\footnote{The table for AME20 was downloaded from %\url{https://www-nds.iaea.org/amdc/}.
%	Extrapolated mass values from this table were ignored.
%}
) combined with three atomic mass tables predicted by theoretical models, namely FRDM92 (\citealt{FRDM95}), FRDM12 (\citealt{FRDM12}), and HFB24 by \cite{Goriely_ea_Bsk22-26}.%
%
%\footnote{The tables for FRDM92, FRDM12 and HFB24 models were downloaded from \url{http://t2.lanl.gov/nis/molleretal/publications/ADNDT-59-1995-185-files.html},
%	\url{http://t2.lanl.gov/nis/molleretal/publications/ADNDT-FRDM2012.html},	
%	and \url{http://www.astro.ulb.ac.be/bruslib/nucdata/hfb24-dat} (\citealt{bruslib}), respectively.}
%
Three representative ash compositions were considered for each of the theoretical mass tables: Superburst (\citealt{KH11}), Kepler (\citealt{Cyburt_ea16}), and Extreme rp ashes (\citealt{Schatz_ea01}}), extracted from \cite{lau_ea18}. As a result, nine models of the outer crust were obtained and $Q_\mathrm{o}(P_\mathrm{oi})$ and $Q(P_\mathrm{oi})$ were calculated for each of them.

Here we model the nuclear reaction chains at the oi interface and in the upper layers of the inner crust. For this purpose we use nuclide abundances for the three ashes obtained in SGC21 as an initial composition at the bottom of the outer crust.
 
To determine the energy density and pressure of free neutrons, we use the BSk24 equation of state proposed by \cite{Goriely_ea_Bsk22-26}. When calculating the total energy density and pressure, we neglect the volume occupied by nuclei (i.e., our microphysical model is the same as in \citealt{CS19_NoEquil}).

Here we present results only for the FRDM12 model. We avoid discussion of the results for FRDM92 and HFB24 models because available mass tables are rather `narrow'. We attempted to extend them with the 31-parameter nuclear mass model of \cite{DZ95} and obtained the results, which are qualitatively the same as for FRDM12. However, quantitatively, they appear to be strongly affected by unphysical effects, associated with merging of the mass tables: the composition of the inner crust is dominated by nuclei near the boundaries of the mass tables, while the stability of these nuclei is associated with spurious energy gap between the merged mass tables.

To account for the nHD condition, we modify the simplified reaction network used for the outer crust in SGC21. The key steps of our simplified reaction network can still be formulated as follows:  (1) checking for allowed reactions in the volume element located in a given crustal layer; (2) performing all allowed reactions by small chunks according to priority rules; (3) compressing the volume element to reach the underlying layer (see \citealt{SC19_MNRAS} and SGC21 for details). However, the details of this procedure are modified.

First, because of the rapid redistribution of neutrons in the inner crust, the considered volume element $V$ should be attached to nuclei. Furthermore, the processes proceeding in a given crustal layer are determined by specifying not only the pressure $P$ but also a certain value of the neutron chemical potential, $\mu_\mathrm{n}$. The thermodynamics of nonequilibrium crust has recently been analyzed in \cite*{GKC_psi21}, where an appropriate thermodynamic potential $\Psi$ was introduced which tends to a minimum at a given $P$ and $\mu_{\rm n}$,
\begin{equation}
	\Psi=E+PV-\mu_n N_b-TS.
\end{equation}
Here $E$ is the energy, $N_b$ is the baryon number in the volume $V$ (containing neutrons and nuclei), while $T$ and $S$ are the temperature and the entropy. It is $\Psi$, not the Gibbs energy (as in the outer crust), that should be minimized in the inner crust. This implies a modification of the first step of the simplified reaction network in the inner crust: only reactions that reduce $\Psi$ are allowed; the corresponding change in $\Psi$ equals the energy released in nuclear reactions. Note that, in our calculations, we work in the limit of vanishing stellar temperature, which simplifies the problem. 

Next, the priority rules applied in the second step are also modified with respect to SGC21. Namely, we allow for rapid capture of several neutrons due to the permanent presence of free neutrons in the inner crust. As a result, the priority rules become as follows: (a) emission/capture of neutrons, (b) electron emission/capture plus emission/capture of neutrons, (c) pycnonuclear fusion. We do not put any restrictions on the number of emitted/captured neutrons for reactions of types (a) and (b). The pycnonuclear reactions are allowed, if their rate exceeds the accretion rate (see \citealt{SC19_MNRAS} for details); the reaction rates were calculated according to \cite{Yakovlev_ea06} with the astrophysical factors taken from \cite{Afanasjev_ea12}.

Last, but not least, we also modified the third step of our simplified reaction network: compression of the volume element. Namely, by compressing matter we increase not only its pressure, as was done in the outer crust but also the neutron chemical potential according to the nHD condition (see GC20):
\begin{equation}
\mathrm d \mu_\mathrm{n}=\frac{\mu_\mathrm{n}}{\epsilon+P} \mathrm dP.
\label{dmun}
\end{equation}

\vspace{-0.5cm}
%%%%%%%%%%%%%%%%%%%%%%%%%%%%%%%%%%%%%%%%%%%%%%%%%%%%%%%%%%%%%%%%%%%%%%%%%%%%%%
\section{Results}\label{res}
%%%%%%%%%%%%%%%%%%%%%%%%%%%%%%%%%%%%%%%%%%%%%%%%%%%%%%%%%%%%%%%%%%%%%%%%%%%%%%

%%%%%%%%%%%%%%%%%%%%%%%%%%%%%%%%%%%%%%%%%%%%%%%%%%%%%%%%%%%%%%%%%%%%%%%%%%%%%%
\begin{figure}
	\includegraphics[width=0.992\columnwidth]{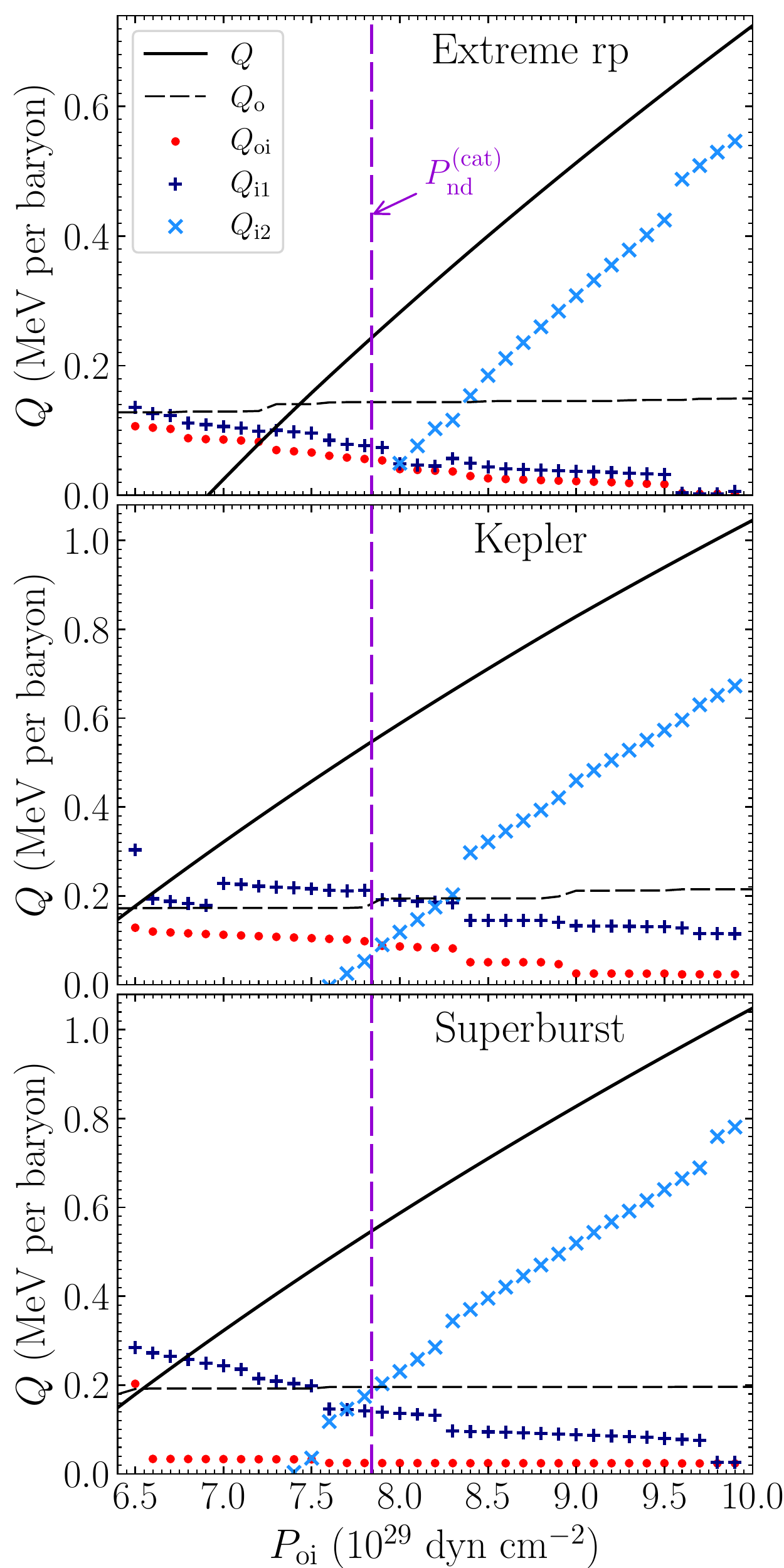}
	\caption{Crustal heating components parametrized by $P_\mathrm{oi}$. 
		The three panels correspond to the three different compositions of ashes considered in the paper. The solid lines demonstrate the total deep crustal heating $Q$, dashed lines show the outer crust heating $Q_\mathrm{o}$. Dots, pluses, and crosses are for $Q_\mathrm{oi}$, $Q_\mathrm{i1}$, and $Q_\mathrm{i2}$, respectively. The vertical dashed lines indicate $P_\mathrm{nd}^\mathrm{(cat)}$ -- the pressure at the oi interface for the catalyzed crust.}
	\label{fig}
\end{figure}
%%%%%%%%%%%%%%%%%%%%%%%%%%%%%%%%%%%%%%%%%%%%%%%%%%%%%%%%%%%%%%%%%%%%%%%%%%%%%%

We consider a set of accreted crust models for a uniform grid of $P_\mathrm{oi}$, located in the region $P_\mathrm{oi}=(6.5-10)\times 10^{29}$~g\,cm$^{-2}$, where the actual $P_\mathrm{oi}$ is expected to be. For each value of $P_\mathrm{oi}$ we apply the simplified reaction network, as described in section \ref{Sec:Approach}, up to the density $\rho=\rho_\mathrm{dc}$ ($P_\mathrm{dc} \approx 2\times 10^{30}$\,dyn\,cm$^{-2}$) so that we calculate the nuclear evolution of matter in the upper layers of the inner crust; the outer crust composition is given by SGC21.

In this way, we explicitly obtain $Q_\mathrm{oi}$, $Q_\mathrm{i1}$, and find $Q_\mathrm{i2}$ from equation (\ref{eq2}). The results are shown in Fig.\ \ref{fig} (a separate panel is presented for each of the three ash compositions). The values of $Q$ and  $Q_\mathrm{o}$ were calculated, as functions of $P_{\rm oi}$, in SGC21 and are also shown in Fig.~\ref{fig} for completeness. For convenience, $P_\mathrm{nd}^\mathrm{(cat)}$ is indicated by vertical line.

Let us describe the general features of nuclear evolution in the inner crust (see subsequent paper for more details). At the bottom of the outer crust, the neutrons are bound in nuclei and the neutron absorption is energetically favourable. Thus, immediately after crossing the oi interface nuclei start to capture neutrons.  The neutron enrichment of nuclei leads to electron emissions (beta-decays), which drives nuclear evolution to the local minimum of the potential $\Psi$. It is worth noting, that the second step -- electron emission -- is important. It leads to substantial dependence of $Q_\mathrm{oi}$ on $P_\mathrm{oi}$ (Fig.\ \ref{fig}). This is due to the fact, that electron emission can be blocked by the growth of the local electron chemical potential as $P_\mathrm{oi}$ increases. 

In the deeper layers of the inner crust, the neutron chemical potential gradually increases (according to equation \ref{dmun}), leading to additional neutron captures followed by electron emissions, although an increase of the electron chemical potential may allow for electron captures for some nuclear species. Pycnonuclear fusion of light elements also contributes to the heating.

As discussed in SGC21, the total deep crustal heating, $Q$, is a monotonic, continuous function of $P_\mathrm{oi}$. However, its components can possess jumps (see Fig.~\ref{fig} and SGC21), associated with thresholds of nuclear reactions (obviously, the most profound jumps are associated with the most abundant nuclei in the crust). The exact positions of such jumps are determined by the nuclear masses (FRDM12 in our case), being rather model-dependent. Moreover, in a more detailed reaction network, these jumps are likely to be smoothed by the thermal effects. That is why, we do not make an effort to identify them precisely and display heating components by symbols, as they were calculated on a uniform grid.

According to Fig.~\ref{fig}, $Q_\mathrm{o}$ only slightly increases in the considered region. Indeed, the initial composition for particular ash stays nearly constant as $P_\mathrm{oi}$ increases. At the same time,
$Q_\mathrm{oi}$ decreases for all ashes due to an increase of the electron chemical potential with $P_{\rm oi}$, which suppresses electron emissions and heating at the oi interface. Similarly, $Q_\mathrm{i1}$ also generally decreases, however in contrast to $Q_\mathrm{o}$ and $Q_\mathrm{oi}$, it demonstrates a non-monotonic behaviour. For example, for Kepler ashes it has a positive jump at $P_\mathrm{oi}\approx 7\times 10^{29}$~dyn\,cm$^{-2}$.

For a given $P_\mathrm{oi}$ the heating components shown in Fig.\ \ref{fig} significantly depend on the composition of ashes. In particular, $Q_\mathrm{oi}$ becomes very small, almost constant, for the Superburst ashes for $P_\mathrm{oi}\geq6.6\times 10^{29}$~dyn\,cm$^{-2}$. In contrast, $Q_\mathrm{oi}$ gradually decreases for Kepler and Extreme rp ashes. However, in the latter case $Q_\mathrm{oi}$ becomes negligible at $P_\mathrm{oi}>9.5\times 10^{29}$~dyn\,cm$^{-2}$. 

For the considered values of $P_\mathrm{oi}$ the general decreasing trend for $Q_\mathrm{oi}$ and $Q_\mathrm{i1}$, combined with the rather small growth of $Q_\mathrm{o}$ leads to the strong increase of $Q_\mathrm{i2}$, which increases faster than the total deep crustal heating, $Q$. Such behaviour should have a profound impact on the models of the thermal evolution of accreting NSs: the models with larger $P_\mathrm{oi}$ will have larger deep crustal heating as a whole, but it will be released in the deeper layers of the crust. Clearly, this should strongly influence crustal temperature profiles during an accretion episode and hence subsequent crustal cooling. It opens up an impressive possibility to constrain $P_\mathrm{oi}$ by confronting cooling models with observations.

 The results shown in Fig.\ \ref{fig} can also be used to set a lower limit on $P_\mathrm{oi}$. To this aim we, following GC21 and SGC21, require the positiveness of the energy release in any crustal region. In comparison to SGC21, where a constraint on $P_\mathrm{oi}$ was deduced from the positiveness of the energy release below the outer crust ($Q_\mathrm{inner}=Q_\mathrm{oi}+Q_\mathrm{i1}+Q_\mathrm{i2}>0$), here we explicitly calculate the components $Q_\mathrm{oi}$ and $Q_\mathrm{i1}$ (positively defined by construction, see Sec.\ \ref{Sec:Reac}) and impose a tighter constraint, $Q_\mathrm{i2}>0$. 

As a consequence, for Superburst and Kepler ashes we got $P_\mathrm{oi}\geq 7.4\times 10^{29}$\,dyn\,cm$^{-2}$  and $P_\mathrm{oi}\geq 7.7\times 10^{29}$\,dyn\,cm$^{-2}$, respectively. For both ashes $P_\mathrm{oi}\leq P_\mathrm{nd}^\mathrm{(cat)}$, in contrast to Extreme rp ashes, for which pressure at the oi interface exceeds $P_\mathrm{nd}^\mathrm{(cat)}$, being bounded by $P_\mathrm{oi}\geq 8.0\times 10^{29}$\,dyn\,cm$^{-2}$. 
Because of the monotonic dependence of $Q(P_\mathrm{oi})$, the deep crustal heating is constrained as $Q\geq 0.43$ MeV and $Q\geq 0.51$\,MeV per accreted baryon for Superburst and Kepler ashes, respectively. As in SGC21,  the lower bound is weaker for Extreme rp ashes ($Q\geq 0.28$ MeV).

%%%%%%%%%%%%%%%%%%%%%%%%%%%%%%%%%%%%%%%%%%%%%%%%%%%%%%%%%%%%%%%%%%%%%%%%%%%%%%%%%%%%%%%%%%%% 
%%%%%%%%%%%%%%%%%%%%%%%%%%%%%%%%%%%%%%%%%%%%%%%%%%%%%%%%%%%%%%%%%%%%%%%%%%%%%%%%%%%%%%%%%%%% 
%\vspace{-0.5cm}
\section{Summary}\label{sum}
%%%%%%%%%%%%%%%%%%%%%%%%%%%%%%%%%%%%%%%%%%%%%%%%%%%%%%%%%%%%%%%%%%%%%%%%%%%%%%%%%%%%%%%%%%%% 
In this Letter, we calculate the energy release at the outer-inner crust interface and in the upper layers of the accreted inner crust for the set of realistic thermonuclear ashes using the thermodynamically consistent approach. Our results are parametrized by the pressure at the outer-inner crust interface, $P_\mathrm{oi}$, which encodes all uncertainties in the physics and nuclear reactions in the innermost layers of the fully accreted crust. 

We consider three families of the models, corresponding to realistic thermonuclear ashes (Superburst, Kepler, and Extreme rp). We start our calculations from the bottom of the outer crust, provided that the composition there has already been found in SGC21 (\citealt{SGC_OC21}). We employ the simplified reaction network of SGC21, which has been modified in order to apply it to the nHD inner crust (see Sec.\ \ref{Sec:Reac}). Our physics input includes the FRDM12 model for nuclear masses and the BSK24 model to describe unbound neutrons. These models allow us to perform calculations up to the density  $\rho \leq 2\times 10^{12}$\,g\,cm$^{-3}$, where the predictions of FRDM12 mass table can still be considered as reasonable (\citealt{lau_ea18,Schatz_ea22}). The resulting heat release is shown in Fig.\ \ref{fig}. The reaction chains and composition profiles are planned to be discussed in the subsequent publication.

The thermodynamic consistency of the nHD approach (GC21, \citealt{GKC_psi21}) allows us to calculate the heat released in the remaining region of the crust ($\rho >2\times 10^{12}$\,g\,cm$^{-3}$), $Q_\mathrm{i2}$. The condition $Q_\mathrm{i2}>0$ imposes a lower limit on $P_\mathrm{oi}$: for the Superburst and Kepler ashes we find $P_\mathrm{oi}\geq (0.94-0.98) P_\mathrm{nd}^\mathrm{(cat)}$, while for Extreme rp ashes $P_\mathrm{oi}\geq 1.02 P_\mathrm{nd}^\mathrm{(cat)}$. These constraints bound the deep crustal heating from below: $Q\geq 0.43-0.51$\,MeV per accreted baryon for Superburst and Kepler ashes, and $Q\geq 0.28$\,MeV per accreted baryon for Extreme rp ashes.

As shown in GC21, SGC21, the heat release $Q$ increases with 
$P_\mathrm{oi}$. Here (see Fig.\ \ref{fig}) we reveal how this heat is distributed in the crust for realistic thermonuclear ashes. As 
in the case of pure $^{56}$Fe ash (GC21), the functions $Q_\mathrm{oi}$ and  $Q_\mathrm{i1}$,  which are crucial ingredients for modelling of the crust cooling of transiently accreting neutron stars, generally decrease with the increasing $P_\mathrm{oi}$. In contrast, the heat released in the deepest layers,  $Q_\mathrm{i2}$, increases with $P_\mathrm{oi}$.
This behaviour should significantly affect the predictions of thermal evolution models for accreting neutron stars, supporting the idea (suggested in GC21 and SGC21) that $P_\mathrm{oi}$ can be constrained by comparing these models with observations. 

%\vspace{-0.5cm}
%%%%%%%%%%%%%%%%%%%%%%%%%%%%%%%%%%%%%%%%%%%%%%%%%%%%%%%
\section*{Acknowledgements}
%%%%%%%%%%%%%%%%%%%%%%%%%%%%%%%%%%%%%%%%%%%%%%%%%%%%%%%
The work of N.~N.\ Shchechilin was supported by the Foundation for the Advancement of Theoretical Physics and Mathematics ``BASIS'' (grant \#20-1-5-79-1).

%\vspace{-0.5cm}
%%%%%%%%%%%%%%%%%%
\section*{DATA AVAILABILITY}
%%%%%%%%%%%%%%%%%%%%%%%%%%%%%%%%%%%%%%%%%%%%%%%%%%%%%%%
Data can be provided by the authors upon a reasonable request.

\label{lastpage}
\end{document}